\documentstyle[12pt]{article} 

\def\bo{{\raise.15ex\hbox{\large$\Box$}}}

\def\dag{^{\dagger}{}}

\def\ordless{{\lower2mm\hbox{$\,\stackrel{\textstyle <}{\sim}\, $}}}
\def\ordmore{{\lower2mm\hbox{$\,\stackrel{\textstyle >}{\sim}\, $}}}

\newtoks\slashfraction
\slashfraction={.13}
\def\slash#1{\setbox0\hbox{$\, #1$}
\setbox0\hbox to \the\slashfraction\wd0{\hss \box0}/\box0}

\def\leftrightarrowfill{$\mathsurround=0pt \mathord\leftarrow \mkern-6mu
        \cleaders\hbox{$\mkern-2mu \mathord- \mkern-2mu$}\hfill
        \mkern-6mu \mathord\rightarrow$}
\def\overleftrightarrow#1{\vbox{\ialign{##\crcr
        \leftrightarrowfill\crcr\noalign{\kern-1pt\nointerlineskip}
        $\hfil\displaystyle{#1}\hfil$\crcr}}}
\def\startarray{\left( \begin{array}}
\def\finarray{\end{array} \right)}
\def\starteq{
\begin{eqnarray}}
\def\fineq{\end{eqnarray}
}

\catcode`@=11
\def\underline#1{\relax\ifmmode\@@underline#1\else
$\@@underline{\hbox{#1}}$\relax\fi}
\catcode`@=12
 
\def\to{\mbox{-}}

\newskip\humongous \humongous=0pt plus 1000pt minus 1000pt

\newif\ifdtup
 
\def\textcite#1{Ref.~{\cite{#1}}}

\def\thefootnote{\fnsymbol{footnote}}
 
\def\author#1#2{{\bf #1} \\ {\em #2}\vspace{5mm}}

\def\sect#1{\bigskip\medskip \goodbreak \noindent {\bf #1}
    \nobreak \medskip}

\def\bold#1{\setbox0=\hbox{$#1$}%
     \kern-.025em\copy0\kern-\wd0
     \kern.05em\copy0\kern-\wd0
     \kern-.025em\raise.0433em\box0 }

\def\title#1#2#3#4#5{\thispagestyle{empty}
        \begin{center} \vspace*{1cm} { \bf #3} \\[.5in] {#4{}}
        \end{center} \vfill \centerline{ ABSTRACT}
   {\nopagebreak \noindent\begin{quotation}\noindent {\small #5}
   \end{quotation}} \vfill {#2} \hfill\begin{tabular}{r} {#1} 
        \end{tabular}  \newpage
        \def\thefootnote{\arabic{footnote}}}
\def\prefer{\section*{}
    \list{[\arabic{enumi}]}{\usecounter{enumi}\settowidth\labelwidth{[000]}
      \leftmargin\labelwidth\advance\leftmargin\labelsep \rightmargin=0pt}
        \small \sfcode`\.=1000\relax}

\def\refer#1{\section*{\large \sc {#1}}
    \list{\arabic{enumi}.}{\usecounter{enumi}\settowidth\labelwidth{[000]}
      \leftmargin\labelwidth\advance\leftmargin\labelsep \rightmargin=0pt}
        \raggedright \small \sfcode`\.=1000\relax}

\def\ReFer#1#2{\section*{\large\sc#1}
    \list{[\arabic{enumi}]}{\usecounter{enumi}\settowidth\labelwidth{#2}
      \leftmargin\labelwidth\advance\leftmargin\labelsep \rightmargin=0pt}
        \raggedright \small \sfcode`\.=1000\relax}

\def\REFER#1#2{\section*{\large\sc#1}
    \list{#2 {enumi}.}{\usecounter{enumi}\settowidth\labelwidth{[000]}
      \leftmargin\labelwidth\advance\leftmargin\labelsep \rightmargin=0pt}
        \raggedright \small \sfcode`\.=1000\relax}

\def\startbib{\vspace{1in}\begin{refer}{References}
\small\frenchspacing\nopagebreak}
\def\endbib{\end{refer} \normalsize \nonfrenchspacing}
\def\startfig{\newpage \centerline{{\sl Figure captions}} \begin{itemize}}

\def\endfig{\end{itemize}}
 
\begin{document}

\newcommand{\nn}{\nonumber}
\newcommand{\be}{\begin{equation}}
\newcommand{\ee}{\end{equation}}
\newcommand{\bea}{\begin{eqnarray}}
\newcommand{\eea}{\end{eqnarray}}  
\title{October 2001} {PC045.1001}
{Charming penguin in nonleptonic $B$ decays \footnotetext{${\dag}$ 
Talk given at the 
{\em International Workshop on QCD: Theory and Experiment}, Martina Franca, 
Italy, 16--20 June 2001}}
{\author{T. N. Pham} {Centre de Physique Th\'eorique, \\
Centre National de la Recherche Scientifique, UMR 7644, \\  
Ecole Polytechnique, 91128 Palaiseau Cedex, France}}    
{In the study of two-body charmless $B$ decays as a mean of looking for
direct CP-violation and measuring the CKM mixing parameters in the
Standard Model, the short-distance penguin contribution 
with its absorptive part generated by charm quark loop
seems capable of producing sufficient $B \rightarrow K\pi$ decays rates, as
obtained in factorization and QCD-improved factorization models. However
there are also long-distance charming penguin contributions which also
give rise to a strong phase due to  the rescattering 
$D^{*}D^{*} \rightarrow K\pi$ . In this talk, I would like
to discuss \cite{Isola} a recent work on the long-distance 
charming penguin as a
a different approach to the calculation of the penguin contributions
in $B \rightarrow K\pi$ decays from charmed meson 
intermediate states. Using  chiral effective Lagrangian for light
and heavy mesons, corrected for hard pion and kaon momenta, we show that
the charming-penguin contributions increase significantly the 
$B \rightarrow K\pi$ decays rates from its short-distance contributions, giving
results in better agreement with experimental data.}

\section{Introduction}
Recent measurements by the CLEO \cite{CLEO}, Babar \cite{Babar} and
Belle \cite{Belle} collaboration give consistent values for the $B \to K\pi$
branching ratios, which are respectively $(18.2^{+4.6}_{-4.0}\pm 1.6)
\times 10^{-6}$,  $(18.2^{+3.3 +1.6}_{-3.0-2.0})
\times 10^{-6}$, $ (13.7^{+5.7+1.9}_{-4.8-1.8})\times 10^{-6}$ for 
$B^{+} \to K^0 \pi^{+} $ and $(17.2^{+2.5}_{-2.4}\pm 1.2)
\times 10^{-6}, (16.7\pm 1.6^{+1.2}_{-1.7})
\times 10^{-6}$, $(19.3^{+3.4+1.5}_{-3.2-0.6})\times 10^{-6}$ 
for $B^0\to K^{+} \pi^{-} $ decays. 
The short-distance contributions to  $B \rightarrow K\pi $ 
decays as given by
the penguin operators without charm quark loop in factorization model
seem  to produce the $B \rightarrow K\pi $ decays rates 
too small compared to the data \cite{Ciuchini1}. A better agreement is
obtained by including the so-called charming penguin contribution
in the effective Wilson 
coefficients \cite{Gerard,Fleischer,Fleischermannel,Deshpande1,Ali}. 
In this way an 
absorptive part of the decay amplitude is generated and the
strong phase from this absorptive part can produce CP violation
in  $B \rightarrow K\pi$ decays \cite{Gerard,Bander}.
This approach seems to produce decay rates in agreement with  data, 
at least qualitatively,  as shown previously 
\cite{Ali,Deandrea,Deshpande,Isola1} , where  the charm
quark loop contribution increases the effective Wilson coefficients 
of the strong penguin operators by about $30\%$,  More
recently charm quark effects  computed by this method have
been obtained in recent works dealing with the validity of  
factorization \cite{Beneke,Du,Muta}. Another approach is to assume
that the charm quark contributions are basically 
long-distance effects essentially
due to  rescattering processes such
as, e.g. $B\rightarrow D D_s\rightarrow K\pi$. These contributions, first
discussed  in \cite{Nardulli}, have been more
recently stressed by \cite{Ciuchini1}, where they are called
charming penguin terms. The situation is  similar to 
the $B_{s} \rightarrow \gamma\gamma$
decay for which  the absorptive part obtained in \cite{Choudhury}
is comparable to the short-distance contribution.  
I would like to discuss here  a recent work \cite{Isola} 
on the charming penguin contributions in $B \rightarrow K\pi$ decays. As details
can be found in this reference, I will present only
the main results  of the work. 
\section{Short and Long distance weak matrix element }
In the standard model, effective Hamiltonian for non-leptonic $B$ decays 
are given by 

\begin{equation}
\kern -0.6cm {\cal H}_{\rm eff} = {G_{F} \over \sqrt{2}}\kern -0.1cm \left[V_{ub}^*
V_{us}(c_1 O_1^u + c_2 O_2^u) + V_{cb}^* V_{cs}(c_1 O_1^c + c_2
O_2^c) \nonumber \\
- V_{tb}^* V_{ts}\left( \sum_{i=3}^{10} c_i O_i + c_g O_g \kern -0.1cm
\right)\right]
\label{Heff}
\end{equation}
where $c_i$ are the Wilson coefficients evaluated at the normalization
scale $\mu = m_b$ \cite{Fleischer,Deshpande1,Buras,Ciuchini,Kramer} and
next-to-leading QCD radiative
corrections are included. $O_1$ and $O_2$ are the usual tree-level
operators, $O_i$ ($i=3,..., 10$) are the penguin operators and
$O_g$ is the chromomagnetic gluon operator. 

The $B \rightarrow K\pi$ decay amplitude $A_{K\pi}$ is given by
\begin{equation}
A_{K \pi}~=~<K(p_{K})\pi(p_{\pi})\vert i{\cal H}_{\rm eff}\vert B(p_{B})> \; .
\end{equation}
In the factorization approximation, the above matrix element is
evaluated at the tree-level as higher order QCD radiative corrections
are already included in the effective Wilson coefficients and the
charm quark operators $O_1^c $ and $O_2^c $ do not contribute. The
short-distance part $A_{SD}$ is obtained with
$c_2=1.105,~c_1=-0.228,~c_3=0.013,~c_4=-0.029,~c_5=0.009,~c_6=-0.033$
\cite{Buras}; $\vert V_{ub}\vert=~0.0038,~V_{us}=~0.22,
~V_{tb}\simeq1,~V_{ts}=-\,0.040$ and 
$\gamma$ = $-\,arg \left(V_{ub}\right)=$
$~54.8^o$\cite{ciuchini3} and $F_0^{B\rightarrow \pi}(m_K^2)\ =\ 0.37$. We find 
\bea
A_{SD}(B^+ \rightarrow K^0 \pi^+) &=& 2.43\times 10^{-8}~{\rm GeV}\cr
A_{SD}(B^0 \rightarrow K^+ \pi^-)&=& \left( 1.86 -i\, 0.95\right) \times
10^{-8}~{\rm GeV}~.
\label{asd1}
\eea
As mentioned, the $B \rightarrow K\pi$ branching ratios obtained 
from Eq.(\ref{asd1}) are too small compared with experiments.  
Instead of using perturbative QCD to treat the charm quark loop
contributions,  
we now  consider the one-particle
 $D,D^{*}$ intermediate state contribution
to the T-product of  two charged weak currents corresponding to the
local operators $O_2^c $. The
matrix element of  $O_2^c $ is evaluated using 
a sum rule due to Wilson \cite{Wilson} . Following Wilson, consider now
the short-distance limit of the T-product of two weak currents
\be
 T\left[J_{\mu N}(x)J_{\mu S}(0) \right ] = B_{1}^{\prime}(x){\sigma}_{m
}^{\prime}(0) 
\ee
where the  contributions from the more singular,
lower dimension operators have been taken out.
$B_{1}^{\prime}(x)$ is the coefficient of
the  local operator ${\sigma}_{m}^{\prime}(0) $. Let
 $M_{AB}(q) = \int d^{4}x \,\exp(iq\cdot x)<A\vert J_{\mu S}(x)J_{\mu
  N}(0)\vert B>  $ we have in momentum space,
\be
\int^{q_{max}}M_{AB}(q)d^{4}q =
B_{1}^{\prime}(q_{max}){\sigma}_{AB}^{\prime},\qquad
B_{1}^{\prime}(q_{max})= \int^{q_{max}}B_{1}^{\prime}(q)d^{4}q
\ee
If $B_{1}^{\prime}(x)$ scales as ${(x^{2})}^{0}$ as in QCD, and for
$q_{max}$ not too large, we obtain
\be
\int^{q_{max}}M_{AB}(q)d^{4}q = {\sigma}_{AB}^{\prime}
\label{mab}
\ee
Eq.(\ref{mab}) thus gives us the matrix element of the 
local operators in terms of a Cottingham-like formula evaluated only
up to a cut-off momentum $q_{max}$ as the high momenta of the integral
has already been factorized in the Wilson coefficients, as stressed in 
previous work \cite{Nardulli,Nardulli1,Pham}. It should be stressed here
that in factorization model, the exchange term in the effective
Hamiltonian is usually Fierz-reordered into a product of two color-singlet 
operators and then evaluated by vacuumm saturation. Actually, it can also
be expressed in terms of an integral over the virtual momentum $q$
which is the difference of the two quark momenta in the initial and final
hadron. For example, the exchange term in the $K\pi$ transition 
is given as ($\psi(k,k-p)$ is the pion B-S wave function),
\be
A(K^{-} \rightarrow \pi^{-}) = \int { d^{4}k \over (2\pi)^{4}}
\int{ d^{4}k^{\prime} \over (2\pi)^{4}} \bar{\psi}(k,k-p)\,
T_{W}(k,k-p;k^{\prime}, k^{\prime}- p^{\prime})\psi(k^{\prime},
k^{\prime}- p^{\prime}) 
\ee 
Making a change of variable $k^{\prime} = q + k$, we have
\be
A(K^{-} \rightarrow \pi^{-}) = \int { d^{4}q \over (2\pi)^{4}}\,T(p,q)
\ee 
\be
T(p,q) = \int { d^{4}k \over (2\pi)^{4}}\bar{\psi}(k,k-p)\,\,T_{W}(k,k
-p;k +q ,k+q -p)\,\psi(k+q,k+q -p)
\ee
which is a higher twist contribution to the  forward virtual 
scattering of the $W$ boson   with momentum $q $ off a hadron. A similar 
expression can
also be given for the transition $\Sigma \rightarrow p$ in hyperon nonleptonic
decays.
The above expression shows that nonleptonic weak matrix elements can
 be expressed as integral over the virtual $W$ boson scattering
amplitude. We have, for the long-distance part $A_{LD}$
\begin{eqnarray}
A_{LD} &=& A_{LD}(B^+ \rightarrow K^0 \pi ^+) ~=~ A_{LD}(B^0 \rightarrow K^+ \pi
^-)~=~ \nn \\ &=& {G_F \over \sqrt{2}}\,V_{cb}^* V_{cs}\, a_2\int
\,{d^4q \over (2\pi)^4}\theta(q^2 + \mu^2)\,T(q,p_B,p_K,p_{\pi})
\label{ald}
\end{eqnarray}
where $\mu$ (or $q_{max}$) is a cut-off momentum separating long-distance and 
short-distance contribution.
$T(q,p_B,p_K,p_{\pi})= g^{\mu\nu}\,T_{\mu\nu}$, with
\begin{equation}
T_{\mu\nu} ~=~i\,\int \,d^{4}x\,\exp (i\, q\cdot x)
<K(p_K)\pi(p_\pi)\vert {\rm T} (J_{\mu}(x)J_{\nu}(0))\vert B(p_B)>
\label{Tmn} 
\end{equation}
$J_{\mu} = \bar{b}\gamma_{\mu}(1 - \gamma_5)c$ and 
$J_{\nu} = \bar{c}\gamma_{\nu}(1 - \gamma _5)s$.
\begin{table}
\begin{center}
\begin{tabular}{|c||c|c|c|}   \hline
\multicolumn{1}{|c||}{$\mu_\ell$} 
&\multicolumn{1}{c|}{$D$}
&\multicolumn{1}{c|}{$D^{*}$}
&\multicolumn{1}{c|}{Total}   
\\ \hline
$0.5$ & $ -4.66\times 10^{-9} $  & $ 1.62 \times 10^{-8} $ & $1.15\times 10^{-8}$ \\
 $0.6 $ & $ -7.77\times 10^{-9}  $  & $ 2.79\times 10^{-8}$ & $2.01\times 10^{-8}$ \\
$0.7$ &$ -1.19\times 10^{-8} $  & $ 4.40 \times 10^{-8} $ & $3.21\times 10^{-8}$\\
\hline
\end{tabular}
\caption{ \small {Numerical values for the real part of $A_{LD}$
in GeV
for $\mu_\ell\ =\ 0.5-0.7$ GeV. First column refers to the $D$,
the second is the $D^*$ contribution.}}
\end{center}
\end{table}

To compute $A_{LD} $ we saturate the $T_{\mu\nu} $ with the
$D,~D^*$ intermediate states. This gives us the usual $D,D^{*}$ pole
term (Born term) for $T(q,p_B,p_K,p_{\pi}) $. 
To compute these pole terms, we use 
heavy quark effective theory and
chiral effective lagrangian  to obtain the $B \rightarrow D,D^{*}$
and $D \rightarrow K\pi $ and $D^{*} \rightarrow K\pi $ semi-leptonic decay form
factors \cite{lee} which appear at each vertex of the  pole
diagrams. $< ( D ,\ D^* ) \vert J^\mu \vert  B >$ is parameterized 
in terms of
the Isgur-Wise function and the matrix elements 
$< K \pi \vert J^\mu \vert D>$   
and $< K \pi \vert J^\mu \vert D^*>$ are computed using
Chiral Effective Lagrangian for semileptonic decays of heavy mesons
to light pseudo-scalar mesons. We  extrapolate the soft meson 
limit to higher momenta  by using  the
full $D^{*}$ propagator in the pole terms (a similar use of the full
$D^*$ propagator to go beyond the soft pion result
has also been given in \cite{Fajfer}) . We also introduce  a form
factor in the strong $DD^{*}\pi$ coupling constant
(a similar approach is used in semileptonic decays
\cite{casalbuoni}). Including this effect, we obtain, for hard pion,
\begin{equation}
G_{D^*D\pi}=\frac{2m_D \,g}{f_\pi}\,F(\vert \vec p_\pi\vert)~,
\label{GGG}
\end{equation}
where $F(\vert \vec p_\pi\vert)$ is 
normalized by $F(0)=1$ which corresponds to the soft pion limit.
($g\approx 0.4$ ).
This form factor can be evaluated by using
the constituent quark model which gives roughly, for 
$\vert\vec  p_\pi\vert \simeq m_B/2$, 
$ F(\vert \vec p_\pi\vert)~=~0.065\pm 0.035~$.

Since the threshold for the $D,D_{s}$ and   
$D,D_{s}^{*}$ production   is below the $B$ meson mass, the $D_{s}$
and $D_{s}^{*}$ pole term for the $D,D^{*} \rightarrow K\pi$ form factors
have an absorptive part. This pole term  is  in fact a rescattering  term
via the Cabibbo-allowed $B \rightarrow D,D_{s}^{*}$ decays followed by the
strong annihilation process
$D,D_{s}^{*}\rightarrow K\pi$ and can be obtained from  the unitarity 
of  the $B\rightarrow K\pi$ decay amplitude. We have 
\begin{eqnarray}
{\rm Disc}\ A_{LD} &=&
 2\, i\, {\rm Im}\, A_{LD}=\, (-2\pi i)^2 \int\frac{d^4
q}{(2\pi)^4}\delta_+(q^2-m^2_{D_s})\, \delta_+(p^2_{D^{(*)}}-m^2_D
)\times \cr &&\cr&\times & A(B\rightarrow D_s^{(*)}D^{(*)})\,
A(D_s^{(*)}D^{(*)}\rightarrow K\pi)\ =\cr&&\cr &=& \, -\frac{m_D}{16\pi^2
m_B}\, {\sqrt{\omega^{*2}-1}}\, \int d\vec n \, A(B\rightarrow
D_s^{(*)}D^{(*)})\, A(D_s^{(*)}D^{(*)}\rightarrow K\pi)\ ,
\end{eqnarray}
With the $A(B\rightarrow D_s D) $,  $A(B\rightarrow D_s^{*} D^{*}) $ given by
factorization and  $A(D_s D\rightarrow K\pi)$, $A(D_s^{*} D^{*}\rightarrow K\pi)$ by
the $t$-channel  $D,D^{*}$ exchange pole terms which are  proportional to 
$G_{D^*D\pi}^{2} $ and could be large due to the factor $m_{D}^{2} $. 
However the rescattering amplitudes $A(D_s^{*} D^{*}\rightarrow K\pi)$ etc.
which are  exclusive processes at high energy, should be
suppressed. This is taken account by  the suppression factor
$F(\vert \vec p_\pi\vert) $ mentioned above. We find, for the absorptive part  
\be
\rm Im \,A_{LD} =  2.34\times 10^{-8}\,\rm GeV
\label{LD}
\ee
of  which $1.45\times 10^{-8}\,\rm GeV $ and  $0.89\times 10^{-8}\,\rm GeV $
are respectively the  $D,D_{s}$ and $D^{*},D_{s}^{*}$ contributions.
To find the real part, we compute  all Feynman diagrams
obtained with the effective Lagrangian for the weak form factors 
and integrate over the virtual current momentum $q$ up to a cut-off
$\mu =m_{b}$. This includes the direct term and the pole terms
which produce the absorptive part. It is possible to choose 
a cut-off momentum by a change
of variable $q=p_B - p_{D^{(*)}}$ to
the  momentum $\ell$ defined by the formula
\begin{equation}
q=p_B - p_{D^{(*)}}\equiv (m_B-m_{D^{(*)}}) v -\ell\ .
\end{equation} 
As discussed in \cite{Isola}, the chiral symmetry breaking scale is
about $1\rm\,GeV$ and the mean charm quark momentum $k$ for the on-shell
$D$ meson is about $300\rm \, MeV$, the virtual momentum $\ell$ should be
below $0.6\rm \, GeV$, hence 
a cut-off $\mu_\ell\approx 0.6~{\rm GeV}~$. The real part is then given
by a Cottingham formula as follows \cite{lee}
\bea
{\rm Re}\ A_{LD}&=&\frac{i}{2\ (2\pi)^3}
\frac{G_{F}}{\sqrt 2}V_{cb}^* V_{cs}\ a_2 
\int_0^{\mu_\ell^2} dL^2                              
\int_{-\sqrt{L^2}}^{+\sqrt{L^2}} dl_0
\sqrt{L^2-l_0^2}\int_{-1}^{1}d \cos(\theta)\ i    \nn \\    
&& \times \left\{\frac{j_D^\mu\ h_{D\, \mu} }{p_D^2 - m_D^2} +
\frac{\sum_{pol}\ j_{D^*}^\mu\ h_{{D^*}\, \mu} }{p_{D^*}^2 -
m_{D^*}^2} \right\}.
\eea
in the above expression, the coupling constant $g$
are corrected by the form factor $F(\vert\vec p_\pi\vert)$. The results 
for the real part are shown in Table 1 for $\mu_\ell = 0.5-0.7$ GeV. 
Our numerical results show that the long-distance charming penguin
contributions to the decays  $B \rightarrow K \pi$ are significant. These
results agree qualitatively  with a phenomenological analysis of
these contributions  given in \cite{Ciuchini1}. In particular, we
found that the absorptive part due to the $D,D_s$ states is
somewhat bigger than that from the  $D^{*},D_s^{*}$ states, but of
the same sign. The real part due to the $D^{*},D_s^{*}$
states is however $3-4$ times bigger and opposite in sign to the
contributions from the $D,D_s$ states. As shown in Table 1, 
the real part and absorptive
part are of the same order of magnitude, at the $10^{-8}\rm\, GeV$
level. The results for the branching ratios are 
\begin{eqnarray}
{\cal B}(B^+ \rightarrow K^0\pi^+)&=&(2.4^{+2.7}_{-1.9})\times10^{-5} \cr
{\cal B}(B^0 \rightarrow K^+\pi^-)&=&(1.5^{+1.8}_{-1.3})\times10^{-5}~.
\label{risultati}
\end{eqnarray}
which agrees with the results from CLEO \cite{CLEO}, Babar \cite{Babar}
and Belle \cite{Belle} mentioned above. The inelastic FSI strong phase 
we get from the absorptive part will produce a CP violation in $B\rightarrow K\pi$ 
decays via the interference with the tree-level terms. We get, for the 
CP-asymmetry between  $B^0\rightarrow K^+\pi^- $ and $\bar B^0\rightarrow K^-\pi^+ $
decay rates : $A_{CP} = 0.21$ for $\gamma=54.8^0$ which is comparable
with recent results from CLEO \cite{cleocp}.
\section{Conclusion}
In conclusion, we believe that the charmed resonance contributions
we found seem to be capable of producing the charming-penguin
terms suggested in \cite{Ciuchini1}  within theoretical errors.
The strong phase generated by the real charm meson intermediate
states would be the essential mechanism for direct CP violation
in charmless $B$ decays as suggested by \cite{Gerard,Bander}. 
Though our estimate of the real
part get uncertainties from the value of the cut-off momentum $\mu_\ell$
due to various form factors, its strength is comparable with the
short-distance part, though not as important as the long-distance
contribution in $K\rightarrow \pi\pi$ decays.

\bigskip

\sect{Acknowledgments}

I would like to thank G. Nardulli and the organizers of the QCD@Work
at Martina Franca for the warm hospitality extended to me at the Conference.

\end{document}